\documentclass[10pt,english,a4paper]{article}
\usepackage{amssymb}
\usepackage{amsmath}
\usepackage{babel}
\usepackage{multicol}
\usepackage{epsfig}

\begin{document}
{\Large The Lorentz-Einstein length contraction is a real effect}\\

{\footnotesize \bf J.-M. L\'evy \\ 
\it Laboratoire de Physique Nucl\'eaire et de Hautes Energies,
CNRS - IN2P3 - Universit\'es Paris VI et Paris VII, Paris.  \\ \it Email: jmlevy@in2p3.fr}\\
\begin{abstract}
An elementary thought experiment is used to show that once the time dilatation effect is established,
there is no way to escape a dual length contraction.
\end{abstract}
\begin{multicols}{2}
\section{Introduction}
It has been argued recently \cite{JField} that the only real effect following from Einstein's
two postulates is time dilatation and that length contraction is but an artifact due to a confusion
between arbitrary settings and physically determined rates of clocks in various inertial frames. We show that this 
contention can be easily rebuted by a basic thought experiment.
\section{Muon story}
The slowing down of a moving clock with respect to its stationnary twin is the most easily established
consequence of Einstein's two postulates founding special relativity. The well known light-clock
thought experiment allows to derive it in a couple of lines (see e.g. \cite{Krane} for a classic or 
\cite{JML1} for a recent text making use of this old device)\\
The reality of the effect can be ascertained by the observation at ground level of muons produced by 
cosmic rays impinging on the atomic nuclei in the upper atmosphere.\\
These muons, which are also abundantly produced in numerous particle physics experiments, are known
to be unstable and to have an exponentially distributed lifetime of mean value 2.2 microseconds in their
rest frame. This corresponds to a distance of 600 meters from their production point at the speed of light. 
Thus a muon produced at an altitude of 30 km would have a probability smaller than $2.10^{-22}$ to reach the ground.
However, muons of cosmic origin are easily shown to be abundant at sea level with the aid of simple apparatus. 
Given the above probability, the sea level observed flux would probably mean a primary cosmic ray flux 
lethal to any form of life on our planet, if that probability were not enormously increased by the slowing down
of the extremely relativistic muon 'clock' in the Earth rest frame. The fact that unstable particles live 
longer when they move with velocities sizable with respect to that of light is routinely checked and used by high 
energy particle physicists.
\section{Dual story}
The very same light-clock used in the preceding section allows to show that a rigid rod measured in a frame 
in which it is moving at a relativistic speed is contracted in the direction of its motion by the same factor (for
a given speed) by which its acoompanying clock is slowed down (see \cite{JML1}).\\
This, however, is one of the effects whose reality is questionned in \cite{JField}; it must be acknowledged that,
contrary to the time dilatation effect, no experiment has ever been performed which checks directly this contraction,
simply because their is no known way at present to accelerate a macrospcopic object at relativistic speeds.\\
However, nothing forbids to imagine observers traveling together with the cosmic muons that we considered above.
The muon and the observer being at rest relative to each other, the muon 'clock' runs normally with the mean lifetime
already quoted. However, many of these muons reach sea level before decaying, in spite of their not being able to 
travel a distance much larger than 600 meters as determined by their accompanying observers. The only possible 
explanation is that the atmosphere is much thiner to them than it is to earth-bound observers. It is reduced to 
something not much thicker than 600 meters because of the Lorentz length contraction.
There is no escaping this conclusion. Length contraction and time dilatation are dual effects which imply
each other. Therefore the contention of \cite{JField} cannot be correct. The magnetic force undergone by a charge 
moving in the vicinity of a current carrying wire is a slightly indirect but quite convincing demonstration of the 
same phenomenon (see \cite{Rindler}) 
 
\begin{footnotesize}

\end{footnotesize}

\end{multicols}


\begin{thebibliography}{2}
\bibitem{JField} J.H. Field, physics/0606101  
\bibitem{Krane} Kenneth Krane, {\it Modern Physics}, John Wiley \& sons 1983 p. 23
\bibitem{JML1} J.-M. Levy, physics/0606103
\bibitem{Rindler} W. Rindler, Relativity special, general and cosmological, Oxford U.P. (2001) p.150-151 
\end{thebibliography}
\end{document}